\documentclass[aps,prb,showpacs,twocolumn,superscriptaddress]{revtex4}

\usepackage{graphicx}
\usepackage{dcolumn}

\begin{document}


\title{Silicon beamsplitter for Fourier transform spectroscopy at far infrared frequencies}

\author{Geoffry Evans}
\email[]{schmadel@physics.umd.edu}
\affiliation{Department of Physics, University of Maryland, College Park,
Maryland 20742 USA}

\author{D. C. Schmadel}
\email[]{schmadel@physics.umd.edu}
\affiliation{Department of Physics, University of Maryland, College Park,
Maryland 20742 USA}

\author{A. B. Sushkov}
\email[]{schmadel@physics.umd.edu}
\affiliation{Department of Physics, University of Maryland, College Park,
Maryland 20742 USA}

\author{H. D. Drew}
\affiliation{Department of Physics, University of Maryland, College Park,
Maryland 20742 USA}
\affiliation{Center of Superconductivity Research, University of Maryland,
College Park, Maryland 20742, USA}


\date{\today}

\begin{abstract}
We report the performance of a silicon wafer beamsplitter for use for low $\Delta\nu>0.3\text{ cm}^{-1}$ resolution Fourier transform spectroscopy at far infrared frequencies.  We characterize the Si beamsplitter by comparing throughput spectra measured with it to those measured with the standard Mylar beamsplitters commonly used in that range.  We find that the throughput of the silicon beamsplitter is substantially greater than that of the Mylar beamsplitters over most of the IR spectrum, and that they are comparable in some limited ranges.  The 2 mm silicon beamsplitter has an etalon spacing of about 0.7 cm$^{-1}$, which interferes with its use for $0.1 {\text{ cm}}^{-1}<\Delta\nu < 0.3 {\text{ cm}} ^{-1}$.  The average efficiency of the Si beamsplitter is 0.37 compared with a maximum efficiency of 0.35 for Mylar.  The Si is particularly more efficient in the 100 to 400 cm$^{-1}$ range because of absorption in Mylar.
\end{abstract}


\maketitle
\section{\label{sec:level1}Introduction}
For Fourier transform spectroscopy in the far IR $\nu< 1000 \text{ cm}^{-1}$ a pellicle Mylar is commonly used for the beamsplitter.  The Mylar thickness is chosen to give an interference enhancement of the beamsplitter efficiency at the frequencies of interest.  Nevertheless, the maximum efficiency is about 35$(\%)$ and absorption in the Mylar further reduces the efficiency over much of the far infrared.  Moreover, the interference zeros of the Mylar etalon generally necessitates the use of several different thickness Mylar beamsplitters to obtain the desired spectral information. A final disadvantage of the Mylar beamsplitter is that the low frequency response, which is where the signal is usually especially weak, has an $\omega^{2}$ fall off below the lowest etalon peak.  While some of these disadvantages are overcome by the use of beamsplitters comprised of a thin film of Si evaporated on a thin Mylar pellicle, there remain the problems of the low frequency response and the Mylar losses.  

For many applications of far IR spectroscopy only modest spectral resolution is required.  Under these conditions a relatively thick Si plate provides a nearly ideal beamsplitter for the entire far infrared up to where CaF2 or KBr can be used for $\nu > 450\text{ cm}^{-1}$.  The Si beamsplitter has an average efficiency of 37$(\%)$ over this entire spectral range.  For a 2 mm thick Si beamsplitter a useful spectral resolution of $\Delta\nu \geq 0.3\text{ cm}^{-1}$ is achievable.  
In this paper we describe the performance of the 2 mm Si beamsplitter as implemented on a Bomen DA-3 FTIR.   It is found to be equal or superior in throughput to the Mylar over the $< 1000\text{ cm}^{-1}$ range and allows spectroscopy over this range without the need to change beamsplitters.

\section{\label{sec:level2}Experimental}
We studied the efficiency of a 2 mm thick silicon beamsplitter over the far infrared frequency range from zero to 1000 cm$^{-1}$ which we compare with that of several Mylar beamsplitters.  The silicon beamsplitter was fabricated from $>100 Ohm cm$ Si and optically polished to with a flatness about 1 micron. Using a combination of sources, detectors and filters appropriate for different ranges, we measured the raw output of the spectrometer under vacuum.  While we have not directly measured the throughput, the ratio of their outputs equals the ratio of their throughputs.  The result is a relative throughput comparison.  We find that the silicon shows greater throughput than the Mylars over most of this range, and is roughly equal in most of the rest of the spectrum.  In this spectral range the throughput of the silicon has no zeroes at 4 cm$^{-1}$ resolution, whereas the Mylars all have broad intervals of zero intensity with widths on the order of 10 cm$^{-1}$.

\begin{figure}
\includegraphics[width=8.6cm, clip=true]{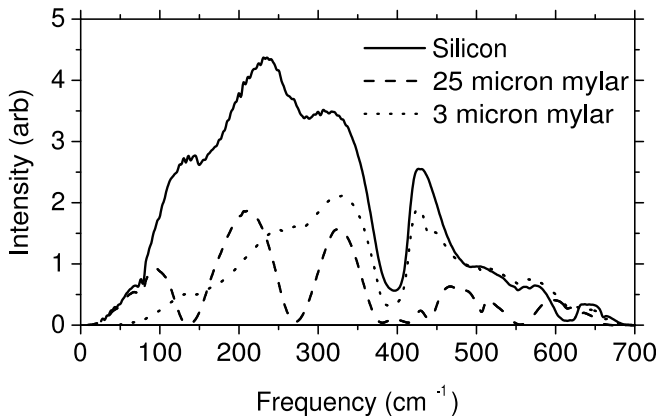}
\caption{\label{fig:thruput_0_to_700} Throughput spectrum of Si beamsplitter compared with that of the $3 \mu$ and $25 \mu$ Mylar beamsplitters with Globar source, 4 K bolometer, at 4 cm$^{-1}$ resolution.}
\end{figure}

\begin{figure}
\includegraphics[width=8.6cm, clip=true]{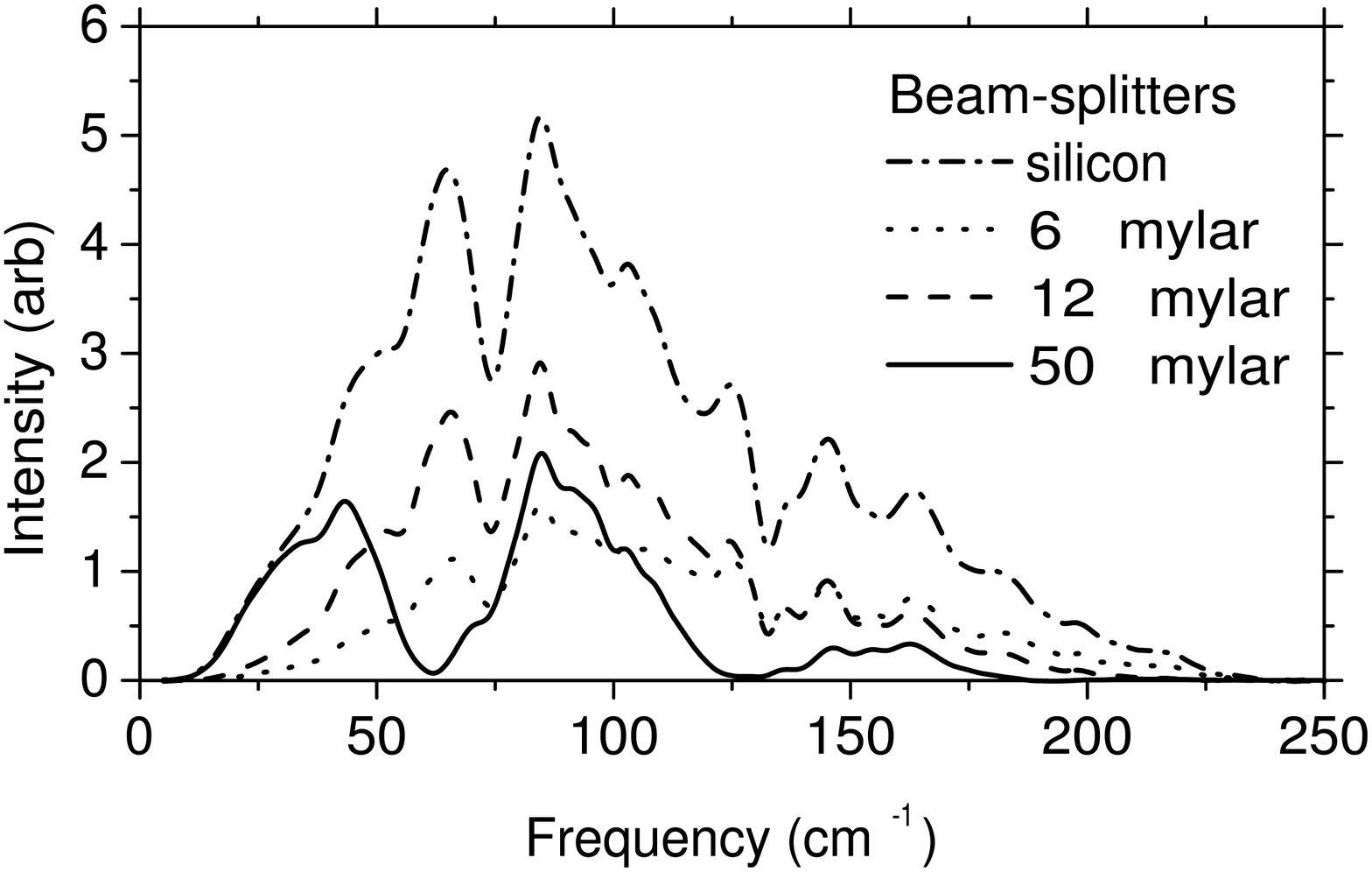}
\caption{\label{fig:thruput_0_to_250} Throughput spectrum of Si beamsplitter compared with 6, 12 and $50 \mu$ Mylar beamsplitter.  Hg vapor lamp, 4 K bolometer, at 4 cm$^{-1}$ resolution.}
\end{figure}

\begin{figure}
\includegraphics[width=8.6cm, clip=true]{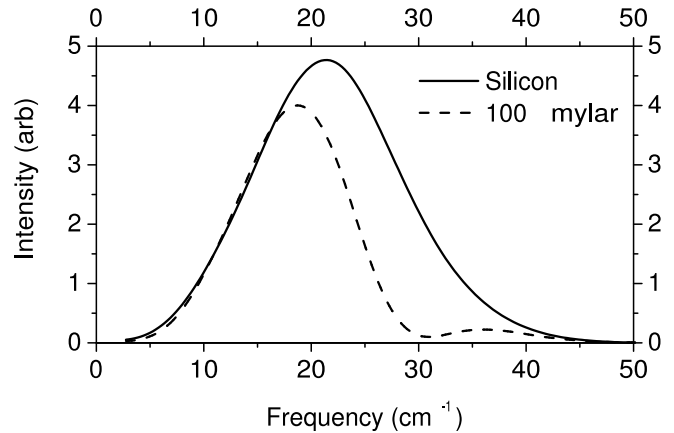}
\caption{\label{fig:thruput_0_to_50} Throughput spectrum of Si beamsplitter compared with $100 \mu$ Mylar beamsplitter with Hg vapor lamp, 4 K bolometer, with 2 mm fluorogold filter at 4 cm$^{-1}$ resolution.}
\end{figure}

Fig.~\ref{fig:thruput_0_to_700} shows the throughput of the silicon along with that of the 3$\mu$ and 25$\mu$ Mylar beamsplitters in the mid infrared.  In each case, we measured the output of the spectrometer with the Globar lamp source and a 4K bolometer detector.  The silicon shows greater throughput than the Mylars over most of this range, and is significantly weaker only in a tiny window between 600 and 640 due to the optical phonon absorption in Si.  Unlike the $25 \mu$ Mylar, the silicon beamsplitter has no broad zeroes in this range.
Fig.~\ref{fig:thruput_0_to_250} Characterization of the throughput of the silicon and the 6, 12, and $50 \mu$ Mylar beamsplitters in the far infrared was made using a mercury lamp source and a 2K bolometer.  We find that the silicon dominates the others from 25 cm$^{-1}$ to 250 cm$^{-1}$ and is roughly equal to them below that.  Again, the silicon has no zeroes at this resolution, while one of the Mylars does.
Fig.~\ref{fig:thruput_0_to_50}shows the throughput of the silicon and the $100 \mu$ beamsplitters below 50 cm$^{-1}$.  In both cases, we measured the output of the spectrometer configured with the mercury lamp source, the 2K bolometer detector and 2 mm thick fluorogold used to filter out radiation above 50 cm$^{-1}$ to minimize baseline errors in the spectrum.  In figure 3 we see that the output of the silicon (and therefore its throughput) is greater over most of this range, and only slightly less between 12 and 17 cm$^{-1}$.  In particular the Si beamsplitter throughput is seen to rise above that of the $100 \mu$ Mylar below 10 cm$^{-1}$.

\section{\label{sec:level}Frequency Resolution}
Like any parallel plate transparent material, both the silicon and Mylar beamsplitters show etalon interference in their transmittance spectra.  Both have periodic bands of zero transmission in their spectra.  In the Mylar beamsplitter the bands are very broad, and give rise to bands of near zero throughput where the beamsplitter can't be used.  

\begin{figure}
\includegraphics[width=8.6cm, clip=true]{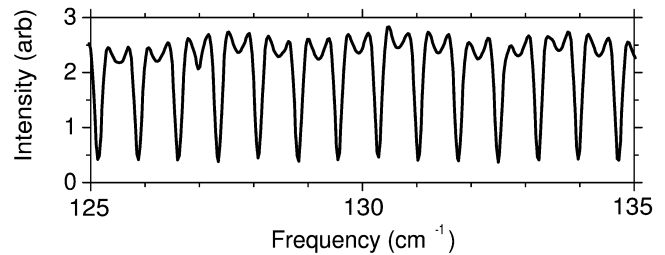}
\caption{\label{fig:rawhires} Throughput spectrum of Si beamsplitter at 0.1 cm$^{-1}$ resolution from 125 to 135 cm$^{-1}$.}
\end{figure}

\begin{figure}
\includegraphics[width=8.6cm, clip=true]{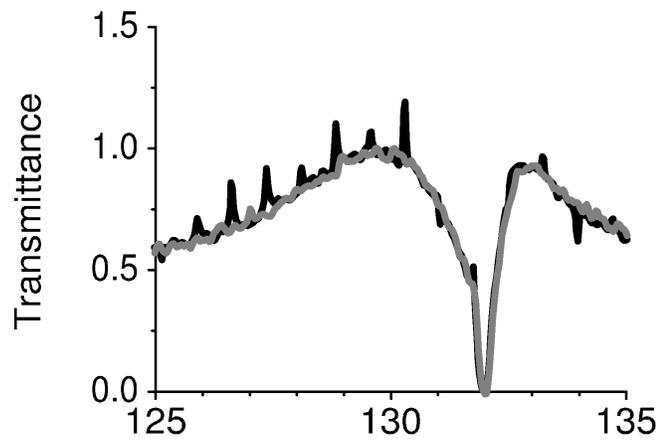}
\caption{\label{fig:rescomparison} Comparison of transmission spectrum of 0.1 mm crystal quartz near the low frequency optical phonon with the Si beamsplitter and $12 \mu$  Mylar beamsplitter at various temperatures.  Resolution was 0.2 cm$^{-1}$.  The spectra were normalized to the transmission of an open hole.}
\end{figure}

In the case of the 2 mm Si beamsplitter the bands are narrow, with a width of about 1 cm$^{-1}$ and an etalon spacing of about 0.7 cm$^{-1}$. Fig.~\ref{fig:rawhires}shows the throughput of the Si beamsplitter for a spectral resolution of 0.1 cm$^{-1}$.  For a resolution $\Delta\nu> 2 \text{cm}^{-1}$ the etalon is averaged and the effective throughput has no zeros.  This allows optical measurements over the whole range of the far IR with a single beamsplitter.  However, for resolution below about 2 cm$^{-1}$ the zeros in the Si etalon have an effect  This is illustrated in Fig.~\ref{fig:rescomparison} where the transmission spectra of a 0.1 mm single crystal quartz sample is shown in the range near the low frequency optical phonon in quartz at 132 cm$^{-1}$ with a resolution of 0.1 cm$^{-1}$.  The width of the phonon was varied by varying the sample temperature down to 5 K.  These spectra were taken with a Globar lamp source and a bolometer, scanning with a resolution of 0.1 cm$^{-1}$.  The transmission data was normalized with the spectrum of an open hole.  The silicon etalon has a period of about 0.7 cm$^{-1}$ but the etalon minima are only about 0.1 cm$^{-1}$ wide.  The Si and Mylar beamsplitters yielded nearly identical spectra apart from sharp noise spikes near the etalon minima of the Si.  From this it is seen that the 2 mm Si beamsplitter can be successfully used for spectral features with widths down to $\Delta\nu \approx 0.3 \text{cm}^{-1}$.  Accurate spectral data for narrower features requires that the position of the feature does not lie within $\sim 0.1 \text{cm}^{-1}$ range of one of the low throughput regions of the beamsplitter etalon.    

We wish to thank Future Instruments Corporation \url{http://www.FutureInstruments.com} for providing the silicon beamsplitter used in the above experiments.



\end{document}